\documentclass[
	a4paper, % Paper size, use either a4paper or letterpaper
	10pt, % Default font size, can also use 11pt or 12pt, although this is not recommended
	twoside, % Two side traditional mode where headers and footers change between odd and even pages, comment this option to make them fixed
]{LTJournalArticle}

\usepackage{graphicx}      % include this line if your document contains figures
\usepackage{amsmath}
\usepackage{amssymb}
\usepackage{bookmark}

\runninghead{} % A shortened article title to appear in the running head, leave this command empty for no running head

\footertext{} % Text to appear in the footer, leave this command empty for no footer text

\title{The LiU-ICE Benchmark \\ An Industrial Fault Diagnosis Case Study} % Article title, use manual lines breaks (\\) to beautify the layout

\author{%
Daniel Jung\thanks{Corresponding author: \href{mailto:daniel.jung@ﬁiu.se}{daniel.jung@liu.se}}, Erik Frisk, and Mattias Krysander
}

\date{\footnotesize Department of Electrical Engineering, Link\"{o}ping University,
Link\"{o}ping, SE-581 83 Sweden}% (e-mail: firstname.lastname@liu.se)}

\renewcommand{\maketitlehookd}{%
	\begin{abstract}
		This paper presents the LiU-ICE fault diagnosis benchmark. The purpose of the benchmark is to support fault diagnosis research by providing data and a model of an industrially relevant system. Data has been collected from an internal combustion engine test bench operated in both nominal and faulty modes. A state-of-the-art model of the air path through an internal combustion engine with unknown parameters is provided. This benchmark has previously been used in a competition at the 12th IFAC Symposium on Fault Detection, Supervision and Safety for Technical Processes (Safe Process) 2024, Ferrara, Italy.  
	\end{abstract}
}

\begin{document}

\maketitle % Output the title section

%----------------------------------------------------------------------------------------
%	ARTICLE CONTENTS
%----------------------------------------------------------------------------------------

\section{Introduction}

An important function in technical systems is fault diagnosis, i.e. the detection of abnormal system behavior and identifying its cause. Diagnosis systems use observations from the system, mainly sensor and actuator signals, and knowledge about the behavior of the system to detect inconsistencies between measurements and model predictions. Fault diagnosis is complicated by model inaccuracies and limited training data. In many applications, faults are rare events meaning that representative data from faults is scarce. Collecting representative faulty data using experiments is expensive, time-consuming, and sometimes not even possible. In addition, developing accurate models is a time-consuming task requiring expert knowledge. 

To address these challenges, this paper presents the LiU-ICE\footnote{Link\"{o}ping University Internal Combustion Engine} fault diagnosis benchmark, an industrial-relevant case study including extensive data collected from an internal combustion engine test bench. Engine fault diagnosis is a non-trivial problem because of non-linear dynamic behavior, and the wide operating range including both transient and stationary operation. The purpose of this benchmark is to support fault diagnosis research by providing a state-of-the-art mathematical model structure describing the system behavior and measurement data from realistic system operation including both nominal and faulty behavior. 

The benchmark model and datasets are provided via gitlab\footnote{\texttt{https://vehsys.gitlab-pages.liu.se/diagnostic\_competition/}}. The model is implemented using the Fault Diagnosis Toolbox which is available in both Matlab and Python \cite{frisk2017toolbox}. The datasets contain sensor and actuator signals from various fault scenarios and fault magnitudes. The fault scenarios include sensor faults and leakages that have been injected during the operation of the system affecting its behavior. This benchmark has previously been used in a competition at the 12th IFAC Symposium on Fault Detection, Supervision and Safety for Technical Processes (Safe Process) 2024, Ferrara, Italy. 

The outline of the paper is as follows. First, the engine case study and test bench are described in Section~\ref{sec:engine}. The provided model of the system is described in Section~\ref{sec:model}. The datasets and fault scenarios are presented in Section~\ref{sec:data}. Some concluding remarks are summarized in Section~\ref{sec:conclusions}
\section{Engine test bench}
\label{sec:engine}

\begin{figure}[!h]
    \includegraphics[width=1\columnwidth]{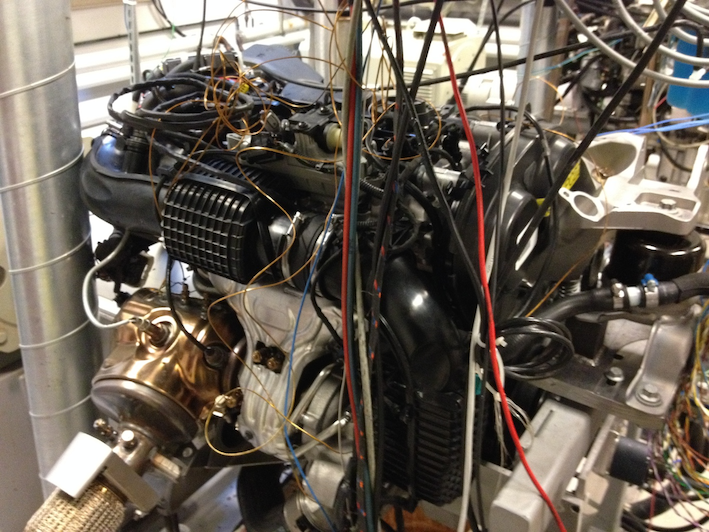}
    \caption{Picture of the engine test bench.}
    \label{fig:engine}
\end{figure}

\begin{figure}[!h]
    \centering
    \includegraphics[width=0.9\columnwidth]{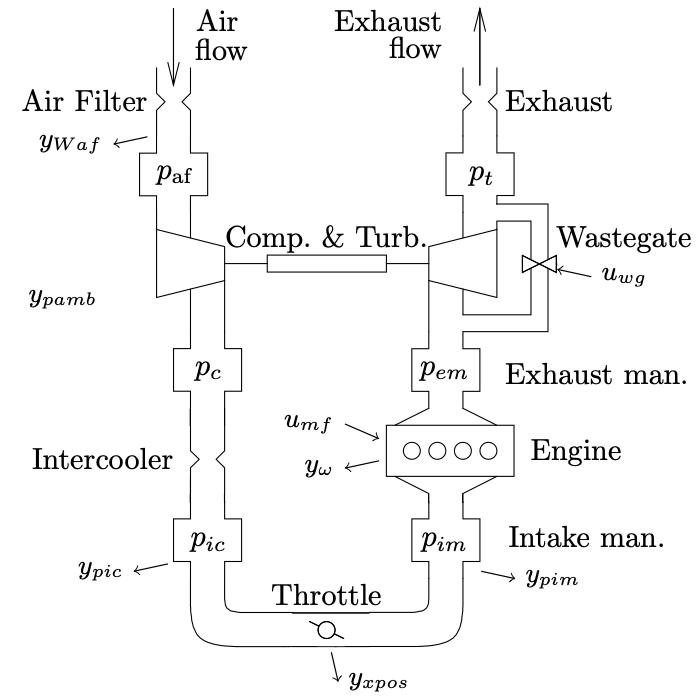}
    %\vspace{-0.1in}
    \caption{Schematic of the air path through the engine (used with permission from~\cite{eriksson2002control}).}
    \label{fig:schematic}
\end{figure}

The case study is a four-cylinder turbocharged internal combustion engine. The engine is mounted on a test bench, as shown in Figure~\ref{fig:engine}. The considered system is the air path through the engine, see Figure~\ref{fig:schematic} where the airflow passes an air filter before the compressor and the intercooler. A throttle is used to control the inlet air to the intake manifold before going into the cylinders where it is mixed with fuel and ignited to generate torque. The exhaust gases pass the exhaust manifold and the turbo that drives the compressor before leaving the exhaust. The wastegate is used to control how much of the exhaust gases that pass the turbo. The engine control unit makes sure that the engine provides the requested torque while controlling the stoichiometry of air and fuel in the cylinder to optimize the combustion and reduce the emissions. 

The available measurements represent a standard setup in a production vehicle including the following eight sensor signals: 
\begin{itemize}
    \item intercooler pressure - $y_{pic}$ 
    \item intake manifold pressure - $y_{pim}$ 
    \item ambient pressure - $y_{pamb}$ 
    \item intercooler temperature - $y_{Tic}$
    \item ambient temperature - $y_{Tamb}$
    \item air mass flow after air filter - $y_{Waf}$ 
    \item engine speed - $y_{\omega}$ 
    \item throttle position - $y_{xpos}$
\end{itemize}    
and the two actuator signals: 
\begin{itemize}
    \item wastegate position - $u_{wg}$ 
    \item injected fuel mass into the cylinders - $u_{mf}$ 
\end{itemize}
 
\section{Model}
\label{sec:model}

\begin{figure}[!h]
    \centering
    \includegraphics[width=1\columnwidth]{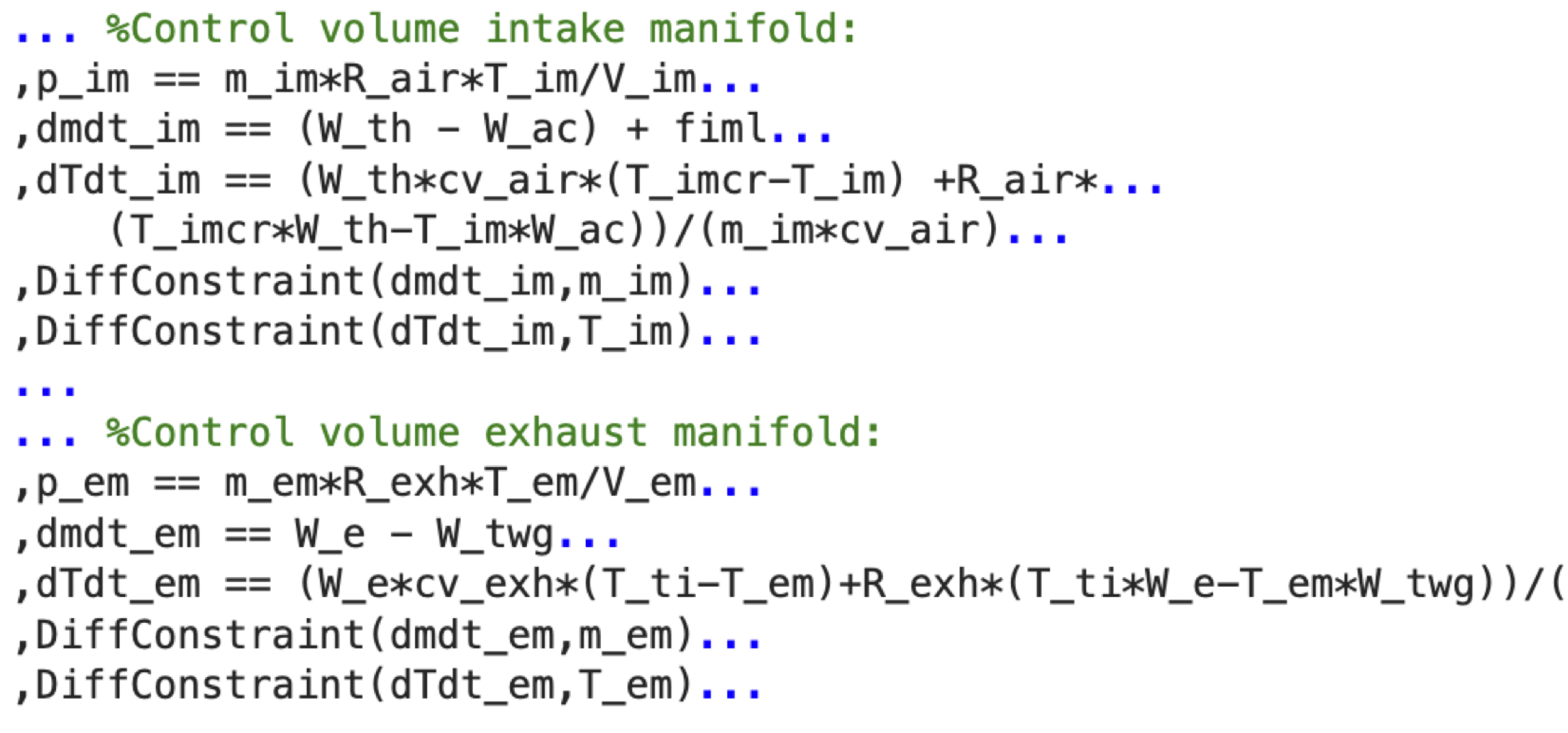}
    \caption{An example of the model implemented in Matlab.}
    \label{fig:matlab_code}
\end{figure}

The structural model is based on a mathematical mean value engine model that has been used in previous works for model-based residual generation, see, e.g., \cite{jung2018combining} and \cite{ng2020realistic}. The mathematical model is similar to the model described in \cite{eriksson2007modeling}, which is based on six control volumes and mass and energy flows given by restrictions, see Figure~\ref{fig:schematic}. The implemented model contains 94 equations (including 14 differential constraints), 90 unknown variables, 10 known variables, and 4 fault variables, and is implemented as a DAE (differential algebraic equations) in the fault diagnosis toolbox \cite{frisk2017toolbox}. An example of the model code in Matlab is shown in Figure~\ref{fig:matlab_code}. However, values for model parameters are not provided.

\begin{figure}[!h]
    \centering
    \includegraphics[width=1\columnwidth]{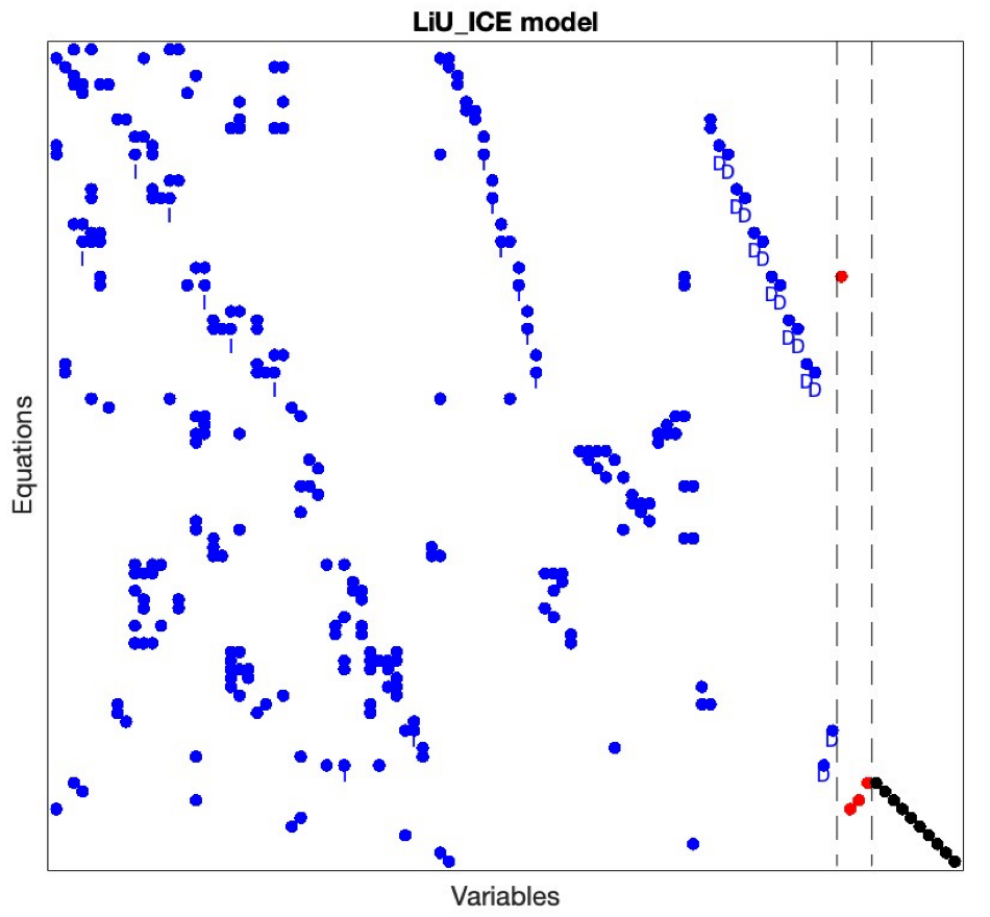}
    \caption{A structural representation of engine model.}
    \label{fig:structural_model}
\end{figure}

The fault diagnosis toolbox contains functionality for analyzing structural diagnosability properties of nonlinear models by analyzing a bipartite graph describing the relation between model equations and variables. Each variable has a unique variable name, including state derivatives. A structural representation of the model is shown in Figure~\ref{fig:structural_model} where each row represents an equation and each column a variable. There is a distinction between unknown variables (blue), fault signals (red), and known variables (black), i.e. sensor and actuator signals. In the structural model, since a state variable and its derivative have different names, additional equations are included in the model to explicitly state the relation between a variable and its derivative. In the toolbox, this is written in the model as 
\begin{equation}
\texttt{DiffConstraint(dx, x)}    
\end{equation}
to show that the model variable \texttt{dx} is the derivative of \texttt{x}. 

\begin{figure}[!h]
    \centering
    \includegraphics[width=1\columnwidth]{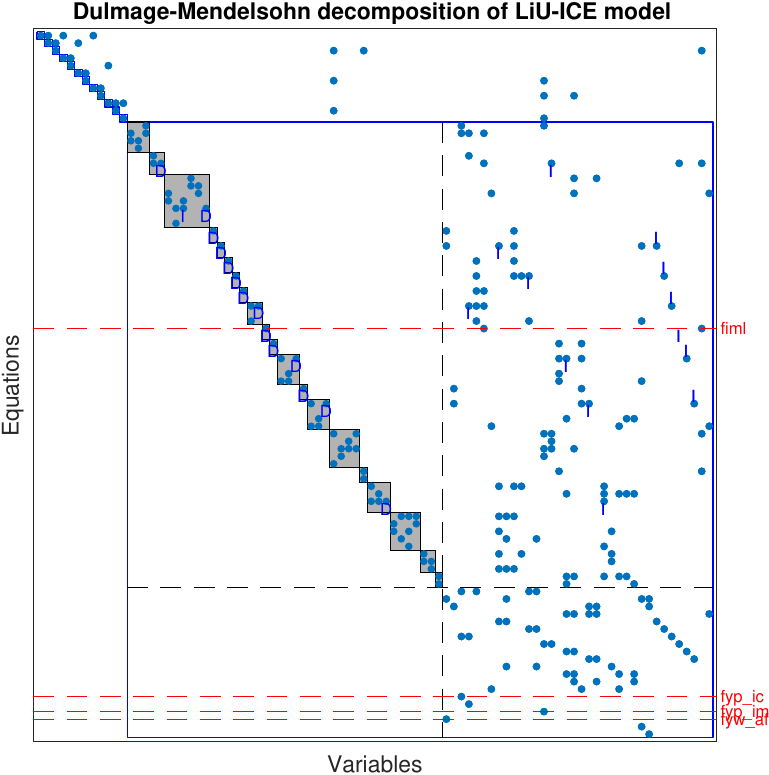}
    \caption{Dulmage-Mendelsohn decomposition of LiU-ICE model.}
    \label{fig:overdetermined}
\end{figure}

A structural analysis of the model in Figure~\ref{fig:structural_model} is shown in Figure~\ref{fig:overdetermined}. The large blue box represents the over-determined part of the model, i.e. the set of equations that can be used to design residual generators. The results using Dulmage-Mendelsohn decomposition show that all faults are %ideally
structurally detectable and isolable from each other \cite{krysander2007efficient}. The degree of redundancy of the model is four. A positive degree of redundancy means that it is possible to find over-determined equation sets that can be used to design residual generators. The number of over-determined equation sets grows exponentially with the degree of redundancy.

\section{Data}
\label{sec:data}

Data is sampled in 20 Hz and stored in CSV format where the first column is the time vector 'time'. Each row contains one sample of data. 

\begin{figure}[!h]
    \includegraphics[width=1\columnwidth]{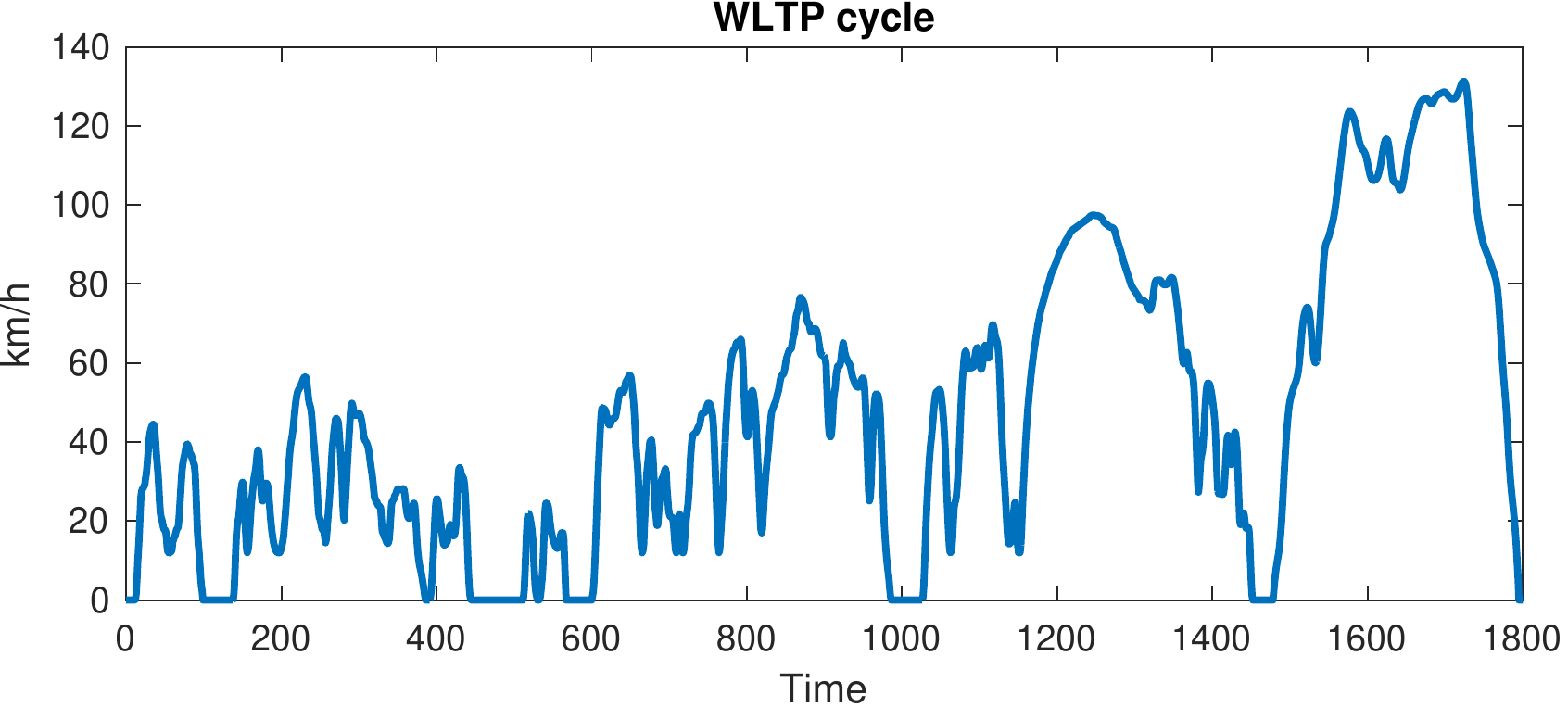}
    \caption{The speed profile of the WLTP cycle.}
    \label{fig:wltp}
\end{figure}

The engine is controlled using a driver model calibrated to follow the WLTP\footnote{Worldwide Harmonised Light Vehicle Test Procedure} cycle, see Figure~\ref{fig:wltp}. The WLTP cycle is approximately 30 minutes long and covers a large operating range representing both urban driving at the beginning of the cycle and highway driving at the end of the cycle.

\begin{figure}[!h]
    \includegraphics[width=1\columnwidth]{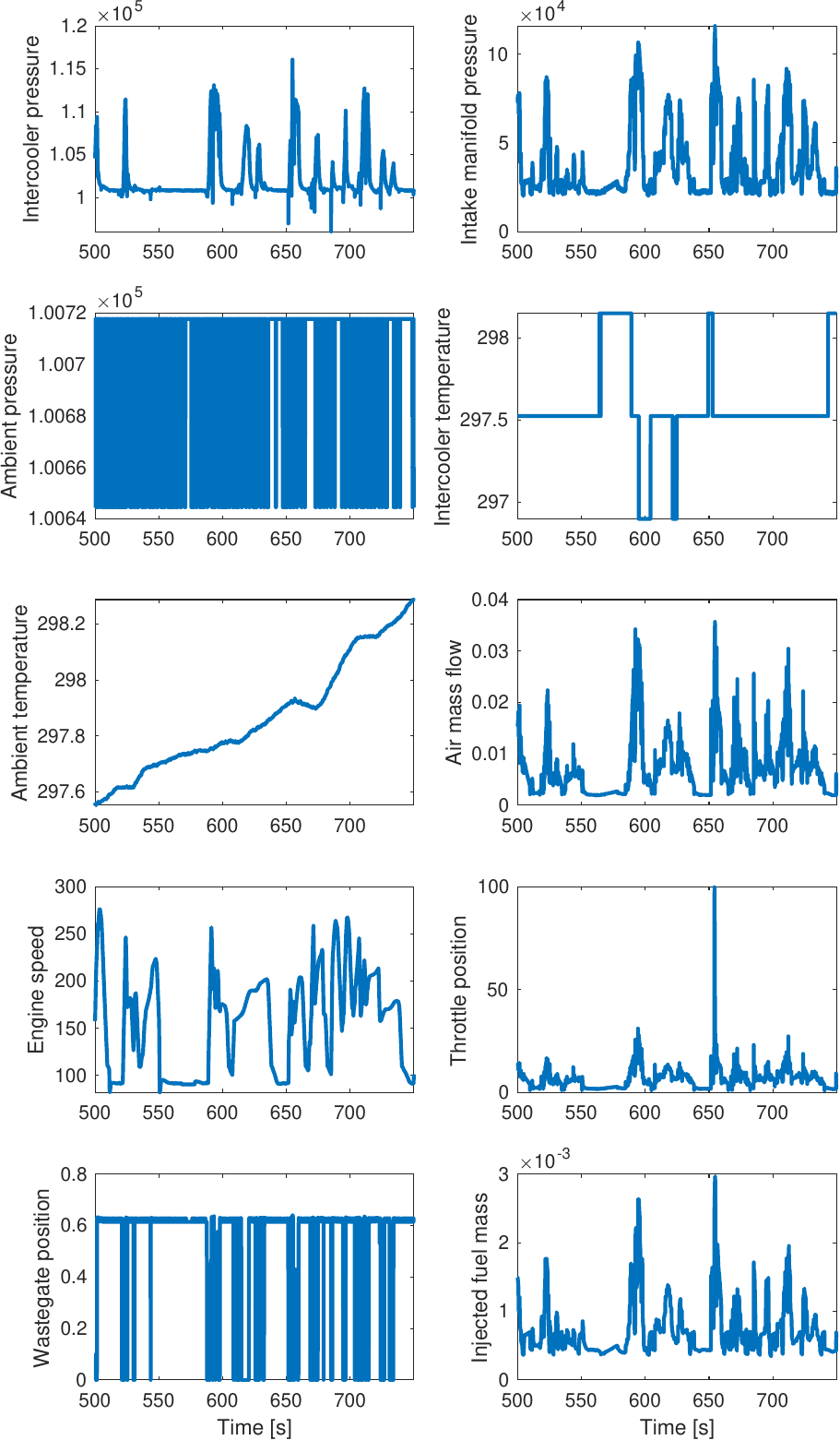}
    \caption{Example of sensor data from the nominal operation.}
    \label{fig:sensor_data}
\end{figure}

Each dataset contains data from one driving cycle and contains one fault scenario including the fault-free mode (NF - No Fault). In the datasets containing faulty behavior, each fault is introduced approximately 120 seconds into each dataset and is present until the end of the scenario. There are in total 25 datasets, 2 fault-free datasets and 23 datasets with different fault scenarios. An example of the provided data is shown in Figure~\ref{fig:sensor_data}.

Data have been collected from four fault modes: 
\begin{itemize}
\item fault in intercooler pressure sensor - $f_{ypic}$
\item fault in intake manifold pressure sensor - $f_{ypim}$
\item fault in air flow mass sensor - $f_{yWaf}$  
\item leakage in intake manifold - $f_{iml}$ 
\end{itemize}

Sensor faults are injected as multiplicative faults $y = (1 + f)x$ directly into the ECU\footnote{Engine Control Unit} where $y$ is the sensor signal, $x$ is the measured quantity, and $f$ is the multiplicative fault signal. Leakages are introduced by manually opening a valve and the size of the orifice is estimated based on the internal diameter of the nozzle. A summary of all fault scenarios is presented in Table~\ref{tb:datasets} where the datasets provided as training data in the competition are highlighted and the rest are provided as test data. 

\begin{table}[hb]
    \caption{Summary of datasets with fault scenarios.}\label{tb:datasets}
\begin{center}
    \begin{tabular}{cc}
        \hline
Fault & Magnitudes \\ \hline
$f_{ypic}$ & -20\%, -15\%,\textbf{-10\%}, -5\%, 5\%, \textbf{10\%}, 15\% \\
$f_{ypim}$ & \textbf{-20\%}, -15\%,\textbf{-10\%}, -5\%, 5\%, 10\%, 15\% \\
$f_{yWaf}$ & -20\%, -15\%,-10\%, -5\%, \textbf{5\%}, \textbf{10\%}, 15\% \\
$f_{iml}$ & 4 mm, \textbf{6 mm} \\
\hline
\end{tabular}
\end{center}
\end{table}

The naming of each dataset is done as follows:
\begin{center}
\texttt{'driving cycle'\_'fault'\_'fault magnitude'.csv}
\end{center}
where 
\begin{itemize}
    \item driving cycle = 'wltp'
    \item fault = 'f\_pic', 'y\_pim','y\_waf','f\_iml'
    \item fault magnitude = For sensor faults the number represents the fault $f$ in $y = (1 + f)x$ where 090 means $(1 + f) = 0.90$, i.e. a $10\%$ reduction in the sensor output, and 110 means $(1 + f) = 1.10$, i.e. a $10\%$ increase in the sensor output. For the leakage fault, 6 mm refers to the orifice diameter in millimeters.
\end{itemize}
\section{Concluding remarks}
\label{sec:conclusions}
The LiU-ICE fault diagnosis benchmark provides a model and data from various faulty scenarios for the fault diagnosis community. The benchmark and proposed setup with training and test data represent an industrially relevant scenario where training data is not representative of all fault scenarios. 

%----------------------------------------------------------------------------------------
%	 REFERENCES
%----------------------------------------------------------------------------------------

\printbibliography % Output the bibliography

%----------------------------------------------------------------------------------------

\end{document}